\documentstyle[12pt,aasms4]{article}
\def \mum{$\mu$m}

\def \msol{M$_{\odot}$}

\newfont{\rten}{cmr10} 
\def\arcdeg{\hbox{$^\circ$}}
\def\arcmin{\hbox{$^\prime$}}
\def\arcsec{\hbox{$^{\prime\prime}$}}
\def \co12{$^{12}$CO}
\def \co13{$^{13}$CO}
\def \kms{km~s$^{-1}$}

\begin{document}
 
\normalsize

\title{A Molecular Counterpart to the Herbig-Haro 1-2 Flow}
\vspace*{0.3cm}
\author{Amaya Moro-Mart{\'{\i}}n \& Jos\'e Cernicharo}
\affil{Instituto de Estructura de
la Materia, Dpto. de Fisica Molecular, CSIC, Serrano 121, E-28006 Madrid,
Spain}
\affil{email:amaya, cerni@astro.iem.csic.es}
\affil{and}
\author{Alberto Noriega-Crespo}
\affil{Infrared Processing and Analysis Center, CalTech-JPL, Pasadena, 
CA 91125, USA}
\affil{email: alberto@ipac.caltech.edu}
\affil{and}
\author{Jes\'us Mart\'{\i}n-Pintado}
\affil{Observatorio Astron\'omico Nacional. Apartado 1143, E-28800 Alcal\'a
de Henares. Spain}
\affil{email: martin@oan.es}
\begin{abstract}

We present high angular resolution (12\arcsec-24\arcsec) and
high sensitivity $^{12}$CO and
$^{13}$CO $J$ = 2-1 and $J$ = 1-0 observations of the HH\,1-2 outflow.
The observations show the molecular counterpart,
moving with a velocity of $\simeq$ 30 km~s$^{-1}$,
of the optical bipolar system driven by the VLA 1 embedded source.
Along the optical jet there
are certain regions where the 
molecular gas reaches deprojected velocities of 100 - 200 km~s$^{-1}$, 
and that we interpret as the molecular jet.
The bipolar CO outflow has a length of $\sim 260\arcsec$ with a 
curved morphology towards the North where it extends
beyond the HH\,1 object ($\simeq$ 120 $\arcsec$) .

Two new molecular outflows have been detected, one arising from
IRAS 05339-0647 which excites the HH\,147 optical flow and another
powered by VLA 2 which drives the HH\,144 optical outflow.
The molecular outflow driven by the VLA 3 source is also
clearly detected and spatially resolved from the VLA 1 main outflow.

\end{abstract}
 

\section{Introduction}

A causality connection between molecular outflows and the optical 
proto-stellar jets is becoming stronger as the number of objects 
with a consistent set of radio, IR and optical observations has grown.
High velocity CO emission is seen to be correlated, kinematically and 
morphologically, for example, with the optical emission in the 
HH 111  outflow (\markcite{cer96}Cernicharo \& Reipurth 1996, hereafter CR96).
In HH 111, high velocity CO bullets are detected 
far away from the optical jet and along the same flow line,
which indicates that the optical jet and molecular outflow coexist spatially
and probably also in time (CR96). Both are manifestations of the
physical processes occurring in the neighborhood of a
nascent star which contribute to dissipate part of its surrounding dense
material.

The HH 1-2 system is one of the brightest Herbig-Haro flows, but
despite various attempts an associated, collimated molecular outflow
has not been clearly identified
(\markcite{mar87}Mart{\'{\i}}n-Pintado \& Cernicharo 1987, hereafter MPC87;
\markcite{tor94}Torrelles et al. 1994;
\markcite{che97}Chernin \& Masson 1995, hereafter CM95;
\markcite{cho97}Choi \& Zhou 1997;
\markcite{cor97}Correia et al. 1997; and see \$3). 
It has been suggested (\markcite{che97}CM95) that
the difficulty in detecting and resolving the CO outflow is
due to the fact that the CO emission is weak and oriented almost in the plane 
of the sky ($\sim 10 \arcdeg$, 
\markcite{nori89}Noriega-Crespo et al. 1989),
and confusion with the background emission is very important.
The source driving the HH 1-2 flow is the embedded object VLA 1
(\markcite{pra85}Pravdo et al. 1985;\markcite{rod90} Rodriguez et al. 1990),
which powers a highly collimated atomic/ionic jet that 
interacts with the surrounding medium exciting the HH 1 and 2 objects.
The flow velocities of the atomic/ionic jet and HH 1-2 are 
$\sim 200 - 380$ \kms~ (\markcite{her81}Herbig \& Jones 1981;\markcite{rag90}
Raga et al. 1990;\markcite{rei93} Reipurth et al. 1993;
\markcite{eis94}Eisl\"offel et al. 1994).
 The similar morphology of the optical and the vibrationally
excited H$_2$ 2.12 \mum~
emission (\markcite{dav94}Davis et al. 1994;\markcite{nori94}
Noriega-Crespo \& Garnavich 1994) indicates a relationship between
the atomic/ionic and the molecular gas dynamics, a conclusion further 
supported by the measured proper motions (\markcite{nori97}Noriega-Crespo et 
al. 1997). It has been suggested (MPC87)
that the distribution of high density gas around HH 1-2
is placed along the walls of the cavity produced by the optical jet
through its interaction with the ambient quiescent gas.
In this Letter we present high resolution and sensitivity maps of the
$^{12}$CO and $^{13}$CO emission which clearly show, 
despite the multiple outflows around the HH 1-2 system, the fast and collimated
CO emission along the stellar jet flow axis.
 
\section{Observations}
 
The observations were performed with the 30m IRAM radio telescope
at different runs during the 1987-1999 period, with a 12\arcsec~
spatial resolution.
The final wide coverage maps were obtained in May 1998
and selected positions along the optical jet in January 1999.
Three SIS receivers at 3, 2 and 1mm were used simultaneously
with receiver temperatures at 100, 100 and 150 K, respectively.
The 3mm and 1mm receivers were tuned to the frequencies of
CO, $^{13}$CO J=1-0 and J=2-1 lines respectively, while the
2mm receiver was used to observe the J=3-2 line of CS (these
data will presented elsewhere).
The spectrometers were two $512\times 1$ MHz filter banks, 
one 256$\times$100 KHz filter bank, and
an autocorrelator with 2048 channels and spectral resolution of 37 KHz.
Calibration was performed by using two absorbers at different temperatures
and by deriving the atmospheric attenuation from the atmospheric
emissivity. In all runs several common positions were observed to
ensure calibration consistency for all data. Intensity differences
never exceeded 10\%.
A region of $316\arcsec \times 440\arcsec$, centered on the VLA 1 source,
was covered. The maps are obtained by fast rastering observations 
with a spacing of 7\arcsec, followed by a pointing and focus check.
Figure 1 shows some representative CO and
$^{13}$CO spectra at selected positions along the optical jet of the
HH 1-2 system (position (0$\arcsec$,0$\arcsec$) corresponds to the VLA 1 
source). The maximum emission of the background cloud is at a velocity of
$\simeq 9.5$ \kms~ in agreement with previous studies
(see, e.g., \markcite{mar87}MPC87; \markcite{cor97}Correia
et al. 1997). Figures 2a and 2b show the integrated intensity of the
$^{12}$CO and $^{13}$CO J=2-1 lines for different velocity ranges.
Finally, Figure 2c shows the superposition of the $^{12}$CO J=2-1
emission, between 11 and 11.5  \kms~(contours) and the [SII] 6717/31 
optical image, covering a field of 2\arcmin.

\section{Results and Discussion}

Previous observations of HH 1-2 in $^{12}$CO were unable to distinguish
the morphology of the outflow arising from the VLA 1 source due to their
limited spatial resolution (\markcite{che97}CM95). Two recent studies 
nevertheless show evidence of a CO outflow along the symmetry axis of 
the optical jet. In the $^{12}$CO $J=3-2$ study of \markcite{cor97}Correia 
et al. (1997), the integrated intensity shows redshifted (V$ = 14-20$ \kms) 
and blueshifted (V $= 0-5$ \kms) emission centered on VLA 1 just as one would
expect for an outflow very close to the plane of the sky. A similar
conclusion was reached by \markcite{cho97}Choi \& Zhou (1997) from
the integrated spectrum within 20\arcsec~around VLA 1 of the same
transition. They detected a blue wing reaching up to V $= -6$ \kms~and red
wing up to V = 28 \kms. The data presented here covers a larger field of
view, with better angular resolution and sensitivity, which permit us to
distinguish the presence of a low radial velocity narrowly collimated 
outflow along the symmetry axis of the atomic jet and the presence of high 
velocity gas at some positions along the jet.
There is a wealth of detail in the $^{12}$CO and $^{13}$CO
maps, so we will focus only on the most relevant aspects in these
data in subsections 3.1 to 3.3. The distribution of other molecular outflows 
in the region is briefly discussed in subsection 3.4.

\subsection{The low velocity gas}

Figure 2a displays the $^{12}$CO emission at different velocity 
ranges in a gray
scale where darker areas mean a higher intensity. These maps show that the
emission between 8 and 14 \kms around the VLA 1 source is narrow.
Figure 2c corresponds to the integrated emission over a velocity
range of 11 - 11.5 \kms, avoiding most of the background cloud emission,
that stresses the highly collimated CO morphology along the optical jet flow, 
bounded by the optical HH 1-2 objects.
The elongated structure seen from 4 to 8 km~s$^{-1}$ north of HH 1,
at around (-40$\arcsec$, 100$\arcsec$), would be the blue-shifted
counterpart of the molecular jet connecting VLA 1 and HH2.
A red-shifted $^{12}$CO intensity enhancement from 10.5 to 14 km~s$^{-1}$
(see Figure 2a) at around (-10\arcsec, 15\arcsec) coincides with the
position and the extension of the optical jet. Indeed the strongest
emission corresponds to the position of a smaller working surface
observed in the H$_2$ 2.12 $\mu$m emission (\markcite{nori97}
Noriega-Crespo et al. 1997). If the CO emission is produced by the interaction
of a working surface with dense material, then the redshifted emission
could be a consequence of the wide velocity dispersion expected
in such structures (\markcite{bohm92}B\"ohm \& Solf 1992).
In addition, the ambient medium around the jet could have a
very axisymmetric distribution. The gas in front of the jet could
contribute with a small column density (the jet is visible), while
the gas behind the optical jet (redshifted gas) could have larger
column densities and densities. A slightly expanding cone around the
jet will be seen mainly at redshifted velocities.

The mass of the HH 1-2 low velocity molecular outflow can be estimated
>from the CO J=1-0 and J=2-1 line intensities. The J=2-1/J=1-0 intensity ratio
(deconvolved for the different beam sizes) shows a clear increase from
1 in the ambient cloud to 1.4-1.7 along the optical jet (i.e. not very
different from that observed in HH 111 (\markcite{cer96}CR96)). If we assume
a thermalised optically thin emission this ratio corresponds to a kinetic
temperature of 30 K. However, due to the low velocity difference between the
CO ambient gas and the molecular outflow, it is difficult to separate their
contributions and to estimate the mass of the molecular outflow. From a map
of the total integrated intensity, which shows a clear enhancement of the CO
emission in the direction of the optical system (see individual panels in
Figure 2a and Figure 2c), we can estimate the CO integrated intensity
associated with the jet by removing the background emission. We obtain an
average value for W(CO(2-1)) of $\simeq$5 K \kms. 
For a distance of 440 pc and by using the W(CO)/Av ratio found by 
\markcite{cer87}Cernicharo \& Gu\'elin (1987)
we derive a mass of 0.1 M$_\odot$ for each lobe of the molecular outflow.
Since this mass is larger than that associated with just high velocity
gas, as e.g. in HH 111 (CR96), we suspect that a low velocity component
is included in our estimates, due perhaps to the walls of 
an expanding cavity around the optical jet (CR96;\markcite{cer97}Cernicharo 
et al. 1997).

The high velocity collimated CO emission associated with the outflow is
observed at a low intensity level and over a narrow velocity range.
It begins to show up at 10.5-11 \kms, but it is only at 11-11.5 \kms~
that it is clearly detected.
This gas has not been observed in previous studies due to their
limited spatial resolution and sensitivity. In HH 111 the cavity walls are 
detected over a velocity range of $\simeq$ 10-15 km~s$^{-1}$ (CR96). 
In HH 1-2 the range is only $\simeq$ 4-5 km/s, but if the projection angle is
closer to its estimated lower limit ($\sim 5\arcdeg$) then both jets have 
similar velocity ranges. We have made some crude estimates
for the volume density and CO column density from a LVG code, 
for the velocity range 11-12.5 \kms~ alone and assuming T$_{K}$ = 30 K, 
we obtain a mass of 0.025 M$_\odot$ in each lobe in HH 1 -2.
Taking into account that the total velocity coverage of the molecular outflow
is $\simeq$6-8 \kms~ the above mass estimates are in good agreement 

\subsection{The high velocity gas}

Although the HH 1-2 system is nearly in the plane of the sky we have
found indications of high velocity gas at several positions along the
optical jet. The top panel of Figure 1 at (48$\arcsec$,-60$\arcsec$),
right on the HH 2 object, displays a large velocity dispersion, as a 
result of the strong shock experienced by the jet in this region. 
At  (12$\arcsec$,-24$\arcsec$) a bright and extended $^{12}$CO red wing is 
clearly detected, reaching up to deprojected velocities of $\sim$ 100 - 200 
\kms (for a 5 - 10$\arcdeg$~angle).
This wing remains in the two following panels, at 
(-12$\arcsec$,12$\arcsec$) and (0$\arcsec$,0$\arcsec$), and belongs to the 
high velocity CO gas in the jet. The jet blue wing is faint but is detected
at (-12$\arcsec$, 12$\arcsec$) and (-12$\arcsec$, 24$\arcsec$), reaching a
deprojected velocity of $\sim 60 - 120 $ kms~$^{-1}$. The high velocity wings
disappear in the last panel at (-36$\arcsec$, 60$\arcsec$) ahead of HH 1.


The high sensitivity data shown in Figure 1 correspond to
selected positions and do not cover the full spatial extent of the jet.
Hence, we can not determine if we are observing continuous high velocity
gas along the jet or discreet bullets as in HH 111 (CR96).
>From the observed $^{12}$CO J=2-1 and J=1-0 intensities we can estimate
the mass of the high velocity gas at each position by assuming a kinetic
temperature for that gas of 60 K, similar to that in other outflows.
We derive masses between 1-5 10$^{-4}$ \msol~ which are typical of
bullets and extremely high velocity gas in young low-mass
star forming regions
(\markcite{bac90}Bachiller et al., 1991; 
\markcite{bac91}Bachiller \& Cernicharo 1990;
\markcite{bac96}Bachiller 1996).
As in the case of HH 111, the momentum and the mass
associated with the high velocity gas seems to be enough to drive
the low velocity gas in the cavity walls.
We have searched for bullets, similar to those found in the HH 111 jet,
further away of the HH objects without success. The more dense ambient
medium in HH 1-2 when compared to HH 111, where the jet and the
bullets are emerging into a very low density gas and escaping from the
cloud, makes the identification of well spatially resolved discrete
emission very difficult.

\subsection{The effect of the jet on the molecular cloud}
At velocities between 2 to 8 km~s$^{-1}$ and at $\simeq 10$ \arcsec~ SE from
HH 2 there is a bright $^{12}$CO emission which could be due to high density
molecular gas struck by the jet. 
A similar intensity enhancement is also detected in the $^{13}$CO 
observations (Figure 2b). The absence of the ambient emission
in \co13 at 9.5 km~s$^{-1}$ and at higher velocities could be interpreted 
precisely as the removal of the molecular gas from this region by the jet.
The overall morphology seen in the $^{13}$CO J=2-1 line is very different
>from the corresponding line of the main isotope. The molecular outflow seen
in CO (see Figures 1 and 2a) is not detected in the $^{13}$CO line.
We detect a blue-shifted `clump' ahead of HH 2 
(see panel 1 of Figure 2b) that seems to be accelerated. This
suggests that the observed optical bow shocks are not the terminal working
surfaces of the outflow (\markcite{ogu95}Ogura 1995), but tracers of a more 
recent outburst event. If so, the clump is not the `cloudlet' detected in 
HCO$^+$(\markcite{dav90}Davis et al. 1990), but rather gas accelerated by the
jet flow. 

Another remarkable feature is the arc shape structure observed in $^{13}$CO
at 10-12 \kms~ arising near VLA 1 (Figure 2b) and that extends $\sim$ 180\arcsec~
SW. This emission has been seen in other molecules (see \markcite{cer91}
Cernicharo 1991) and it could be tracing low velocity gas from a cavity seen
edge-on. The arc disappears at higher velocities (except around VLA 1), and
its overall velocity field does not seem to be that of a rotating toroid,
as suggested by \markcite{marc88} Marcaide et al. (1988) and \markcite{tor94}
Torrelles et al. (1994).
There is an off-center filamentary structure nearly parallel to the atomic jet 
axis and slightly red-shifted (see panel 3 of Figure 2b), which extends
beyond the optical knots of HH 1, tracing perhaps the interaction of the
outflow with the environment (i.e, the cavities discussed by MPC87).

\subsection{Other Outflows around VLA 1}
In addition to the molecular gas associated with HH 1-2 
discussed above, our data also indicate the presence of several molecular
outflows in the region. Previous studies (\markcite{che95}CM95) were not able
to separate the outflow from VLA 1 from the brighter and more massive VLA 3 
outflow. The higher resolution of our data makes this distinction possible,
as it is clearly shown in Figure 2a. The blue lobe, between 2 and 4
km~s$^{-1}$, is found north of VLA 3, while the red lobe, from 12 to 16
km~s$^{-1}$, is south. The  VLA 3 source is
located at (-60$\arcsec$, 45$\arcsec$) with respect to VLA 1. The velocities 
are consistent with the ones derived by \markcite{che95}CM95, 0-4.5 and 12-18
km~s$^{-1}$ for the blue and red lobe respectively.
The $^{12}$CO J=2-1 observations between 10 and 16 km~s$^{-1}$ show an
extended conical structure, with the apex close to the position of 
the near infrared IRAS 05339-0647 source $\sim 1\arcmin$ NE of 
HH 1 and at 2 \arcmin~south of V380 Ori. Two HH objects are also found near
this source, HH 147 A-B. The proper motions derived by
\markcite{eis94}Eisl\"offel et al. (1994) show that the HH 147 outflow points 
towards the SW at a position angle of 230\arcdeg, and away from 
IRAS 05339-0647. The maps in Figure 2a with
velocities between 4 and 8 km~s$^{-1}$ present some $^{12}$CO J=2-1
emission engulfing the region around the optical HH 147 A-B objects in an 
elongated structure oriented SW and could correspond
to the molecular counterpart.
In Figure 2c there is also some $^{12}$CO emission along the HH 144 optical
flow (\markcite{rei93}Reipurth et al. 1993) and it probably represents 
the molecular
counterpart to this outflow arising from VLA 2 and oriented E-W.
This $^{12}$CO emission outside the HH1-2 main axis
is seen at velocities ranging from 10.5 to 12 km~s$^{-1}$ in Figure 2a.

\section{Conclusions}
In the HH 1-2 outflow the bulk of the $^{12}$CO gas ($\sim 0.1 M_\odot$)
is moving at deprojected velocities
of 20-30 km~s$^{-1}$. It is very likely that some of the $^{12}$CO emission
arises from the cavity walls excavated by the jet, but there must be 
a contribution from the molecular jet itself. The presence of clear
traces ($\sim 10^{-4} M_\odot$) of high velocity molecular gas moving
at 100-200 km~s$^{-1}$ reinforces this idea.
There are few examples of bipolar outflows where it
has been possible to observe simultaneously the high velocity molecular gas
associated with the highly supersonic atomic/ionic gas.
The HH 111 outflow (CR96) and now the HH 1-2 outflow are systems where the 
optical jet, the warm collimated H$_2$ gas and the high velocity $^{12}$CO gas 
are coexisting.
The fact that most of the known optical jets are in the plane of the
sky makes it difficult to detect their associated molecular jets.
The present results and those obtained by CR96 suggest that high velocity
molecular gas may be present in other optical outflows, perhaps even
in the form of ``bullets''. In addition, associated to this
gas there is a low velocity molecular component along the cavity walls
excavated by the supersonic atomic/ionic jet.
                                                
\noindent
{\bf acknowledgements :}
We thank the referee, Bo Reipurth, for his critical and careful reading
of the manuscript.
We thank Spanish DGES for support under grants PB96-0883, PNIE98-1351,
and PB96-104.


\clearpage

\begin{center}
Figure Captions
\end{center}

\figcaption[fig1.ps]{Antenna temperature
(in K) as a function of velocity (in \kms), at selected positions 
along the optical jet of the HH 1-2 system. In each panel the upper plot
corresponds to the $^{12}$CO emission at 1 MHz resolution, shifted and
enlarged by a factor 10 in intensity to show the high velocity wings.
The middle line is the $^{12}$CO emission at 100 KHz resolution, and the
histogram is the $^{13}$CO emission at 100 KHz.\label{fig1}}
  
\figcaption[fig2a.ps]
{{\bf (a)} A series of $^{12}$CO J=2-1 maps at different
velocity channels of the HH 1-2 outflow, covering a region of 
$316\arcsec \times 440\arcsec$. The optical knots are overlaid in black 
and labels for some of the sources and the HH object present in the region
are included. The upper grey scale limits are different for each frame, in 
order of increasing velocity: 10, 20, 50, 75, 15, 10, 10, 20, and 20.   
The contours for the maps are: 
>from 1 to 3 by 1; 6 to 14 by 2; 25 to 35 by 2; 36 to 51 by 3; 3 to 8 by 1;
1.5 to 6.5 by 1; 1 to 5 by 1; 3 to 12 by 2; 1.5 to 6.0 by 1
(K \kms). {\bf (b)} $^{13}$CO emission at selected velocity ranges.
The contours are the following: from 
5 to 22.5 by 2.5; 7 to 27 by 2.5; 2 to 12 by 2 (K \kms).
{\bf (c)} A superposition of the $^{12}$CO $J=2-1$
and the optical [S~II] 6717/31 emissions from Reipurth et al. 
(1993) within a 2\arcmin~field of the HH 1$-$2 system. The contours are 
logarithmic and for a velocity range of 11-11.5 \kms, the ambient
gas is at a velocity of 9.5 \kms.
\label{fig2}}

\clearpage

\begin{figure}[h]
\plotone{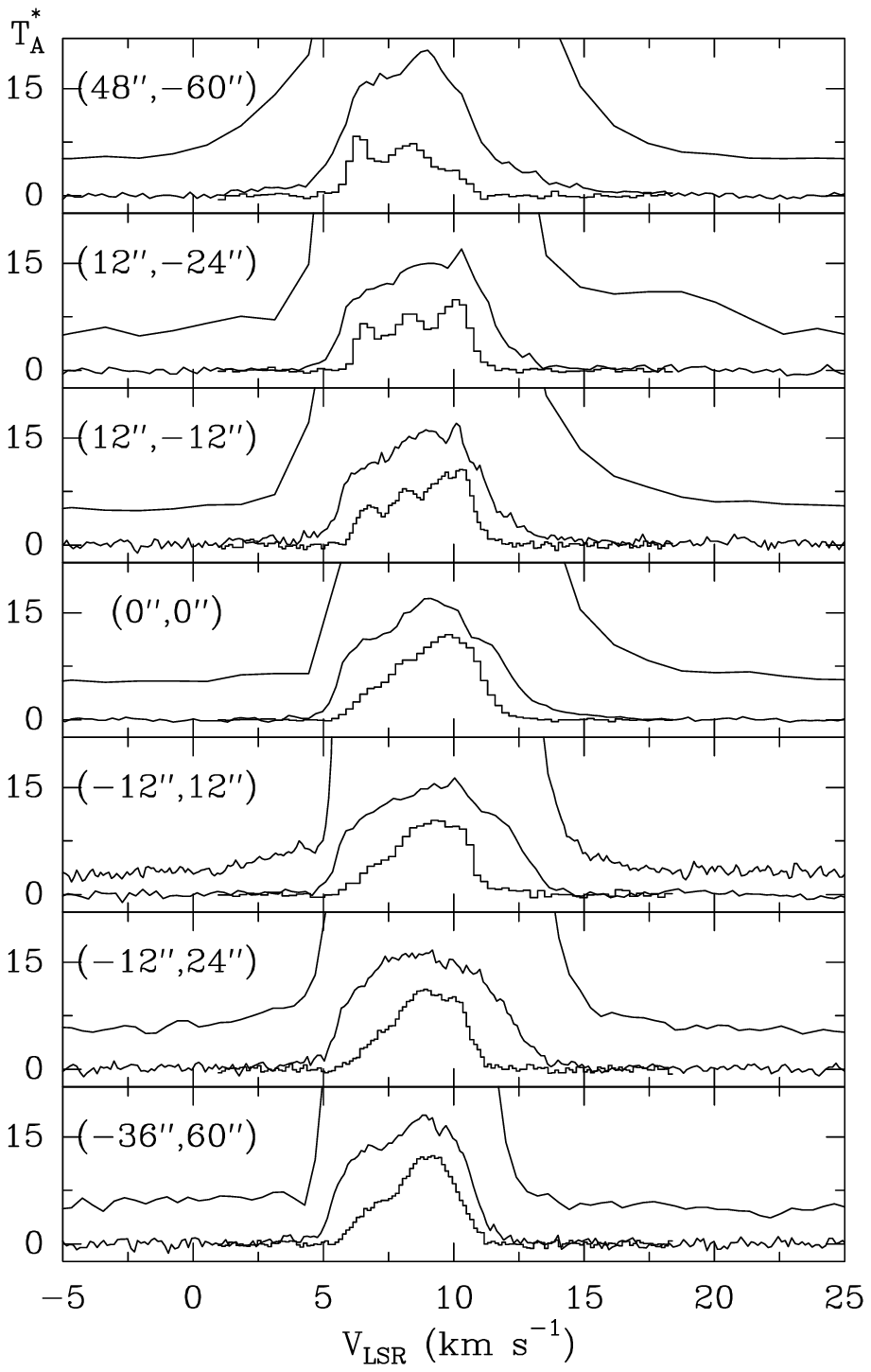}
\end{figure}

\clearpage
 
\begin{figure}[h]
\plotone{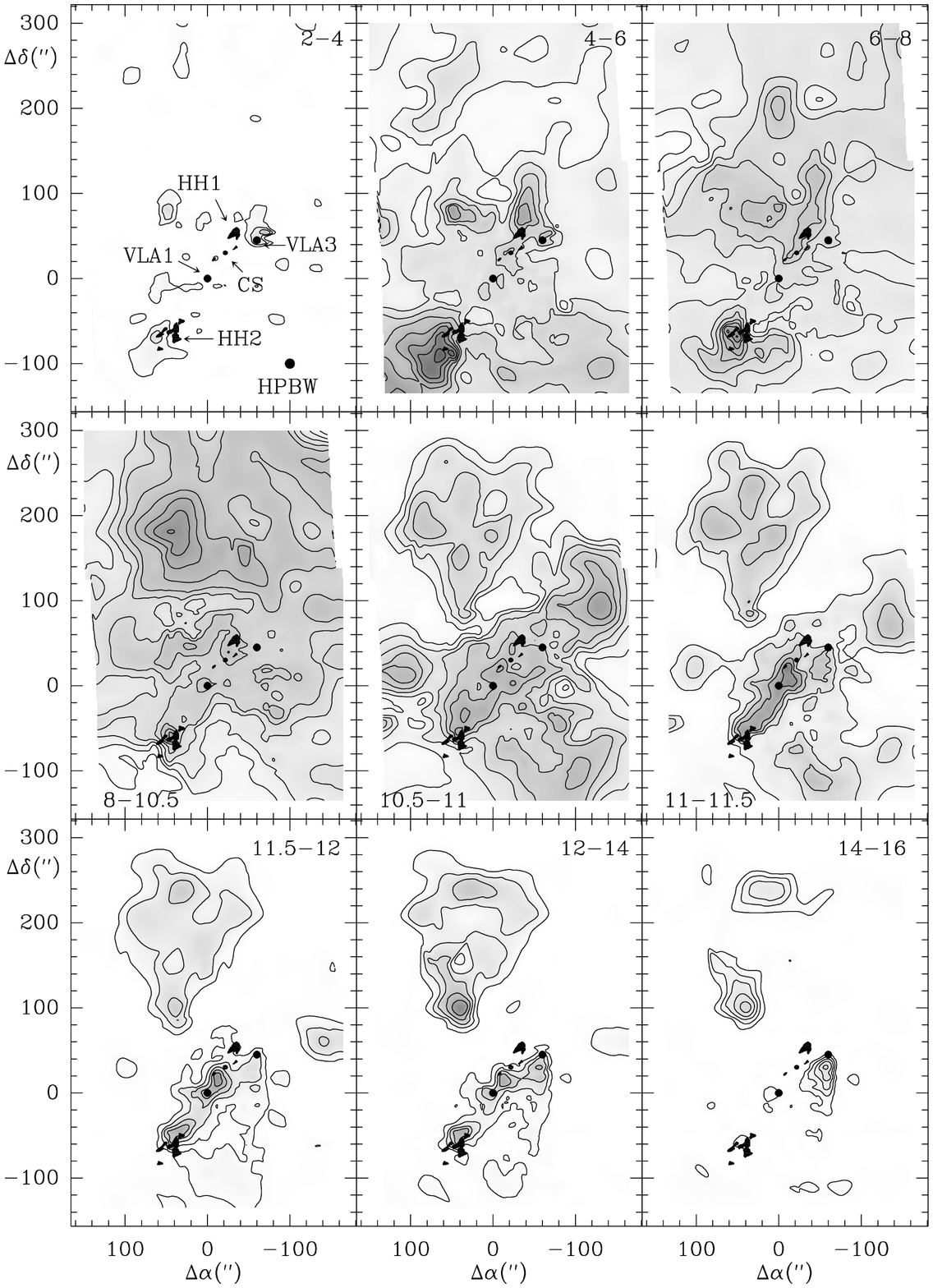}
\end{figure}

\clearpage
 
\begin{figure}[h]
\plotone{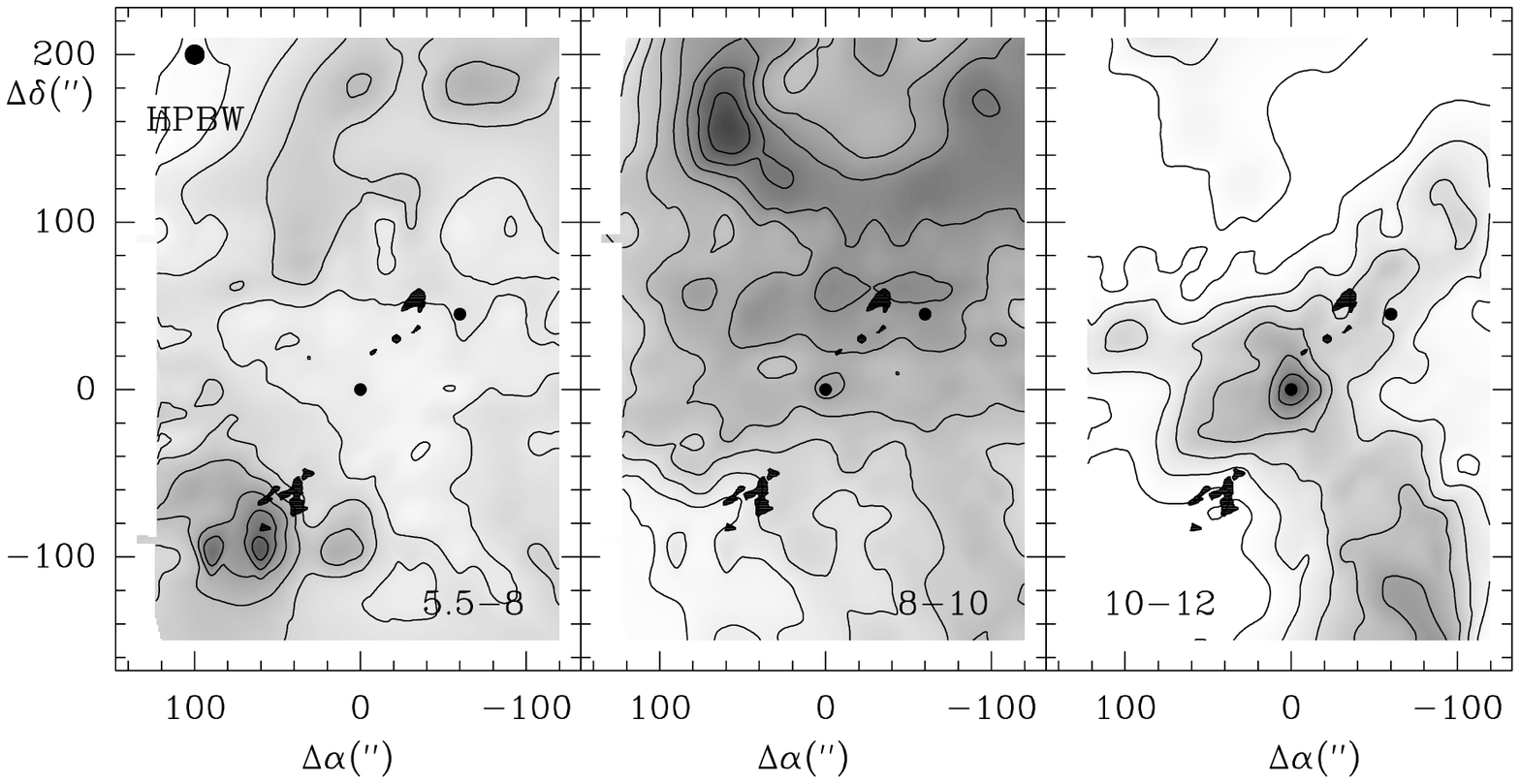}
\end{figure}

\begin{figure}[h]
\plotone{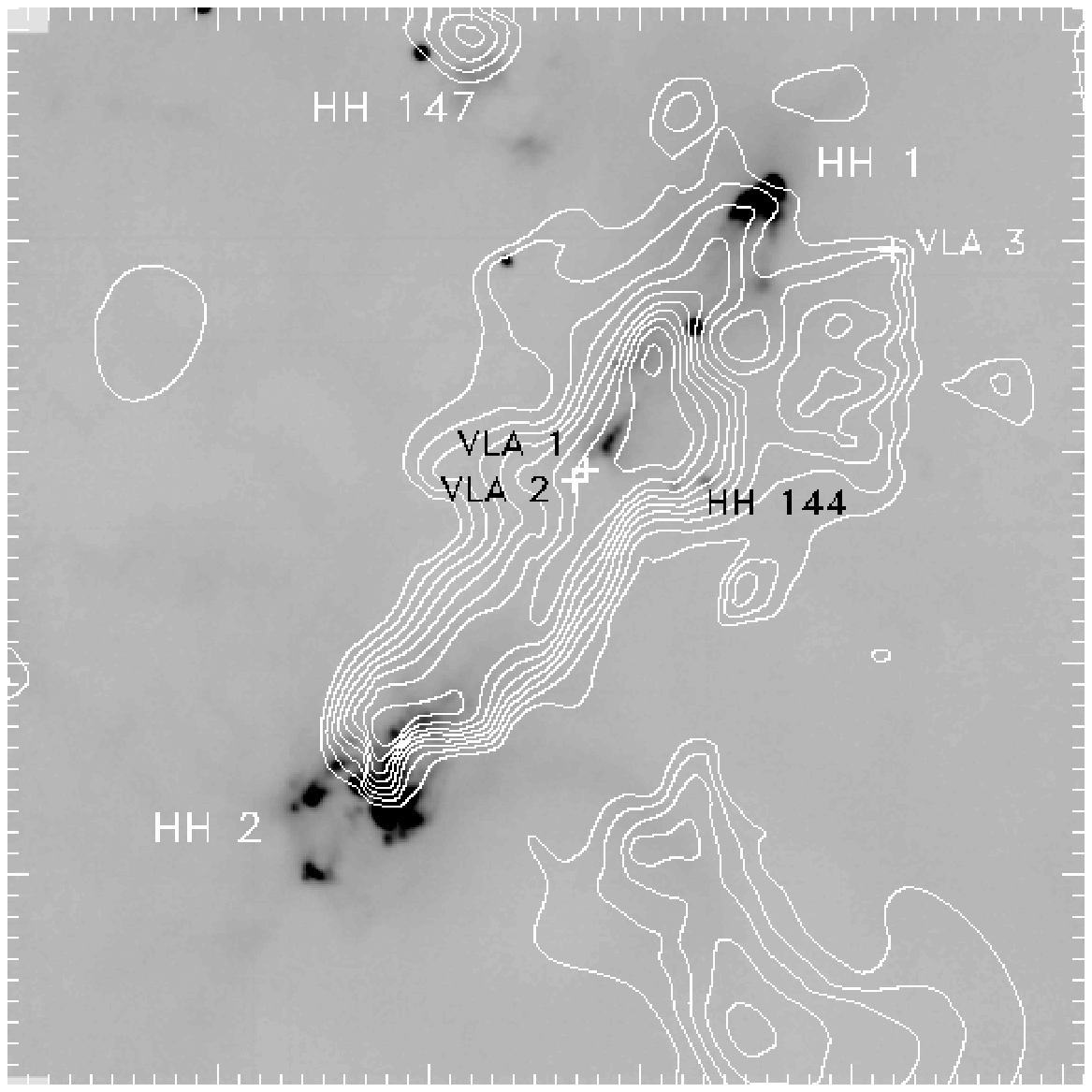}
\end{figure}

\end{document}